\documentclass[]{aastex631}
\usepackage{mathtools}

\usepackage{booktabs} 

\begin{document}

\title{A Statistical Analysis of Magnetic Parameters in Solar Source Regions of Halo-CMEs with and without SEP events}

\author[0009-0009-7015-0024]{Xuchun Duan}
\affiliation{State Key Laboratory of Solar Activity and Space Weather, National Astronomical Observatories, Chinese Academy of Science, Beijing
100101, People's Republic of China}
\affiliation{School of Astronomy and Space Science,
University of Chinese Academy of Sciences,
Beijing 100049, People's Republic of China}

\author[0000-0001-6655-1743]{Ting Li}
\affiliation{State Key Laboratory of Solar Activity and Space Weather, National Astronomical Observatories, Chinese Academy of Science, Beijing
100101, People's Republic of China}
\affiliation{School of Astronomy and Space Science,
University of Chinese Academy of Sciences,
Beijing 100049, People's Republic of China}

\author[0000-0002-9534-1638]{Yijun Hou}
\affiliation{State Key Laboratory of Solar Activity and Space Weather, National Astronomical Observatories, Chinese Academy of Science, Beijing
100101, People's Republic of China}
\affiliation{School of Astronomy and Space Science,
University of Chinese Academy of Sciences,
Beijing 100049, People's Republic of China}

\author[0000-0002-4581-9242]{Yang Wang}
\affiliation{School of Science, Harbin Institute of Technology, Shenzhen 518055, People's Republic of China}
\author{Yue Li}
\affiliation{School of Science, Harbin Institute of Technology, Shenzhen 518055, People's Republic of China}

\author[0000-0001-5933-5794]{Yining Zhang}
\affiliation{State Key Laboratory of Solar Activity and Space Weather, National Astronomical Observatories, Chinese Academy of Science, Beijing
100101, People's Republic of China}
\affiliation{School of Astronomy and Space Science,
University of Chinese Academy of Sciences,
Beijing 100049, People's Republic of China}

\author[0000-0001-5657-7587]{Zheng Sun}
\affiliation{School of Earth and Space Sciences, Peking University, Beijing 100871,
People's Republic of China}
\affiliation{Leibniz Institute for Astrophysics Potsdam, An der Sternwarte 16,
14482 Potsdam, Germany}

\author{Guiping Zhou}
\affiliation{State Key Laboratory of Solar Activity and Space Weather, National Astronomical Observatories, Chinese Academy of Science, Beijing
100101, People's Republic of China}
\affiliation{School of Astronomy and Space Science,
University of Chinese Academy of Sciences,
Beijing 100049, People's Republic of China}

\correspondingauthor{Ting Li}
\email{liting@nao.cas.cn}
\correspondingauthor{Yijun Hou}
\email{yijunhou@nao.cas.cn}

\begin{abstract}
Large SEPs can cause adverse space weather hazard to humans technology and such events are especially associated with halo coronal mass ejections (CMEs). But in turn, a significant portion of halo-CMEs are not associated with large SEPs. The objective of this study is to gain an understanding of the source region distinctions between halo-CMEs in SEP and No-SEP events. Among the 176 halo-CMEs observed from 2010-2024, we screen out 45 large SEP events and 131 No-SEP events from this dataset. It is revealed that CME speed is a good discriminator between SEP and No-SEP events. Through classifying the source regions of all the halo-CMEs, we find that 53\% of SEP events originate from ``Single AR'', and 47\% from ``Multiple ARs'' or ``Outside of ARs''. The corresponding proportion for No-SEP events is 70\% and 30\%. This suggests that SEP source regions are more likely to originate from large-scale sources. We have also calculated the relevant magnetic parameters of the source regions and found that SEP source regions have higher magnetic free energy and reconnection flux compared to No-SEP source regions. However, SEP source regions are smaller in terms of the intensive magnetic parameters such as mean characteristic magnetic twist $\alpha$ and mean shear angles. Our statistical results can provide new potential variables for forecasting SEPs.
\end{abstract}

\keywords{Solar energetic particles (1491); Solar activity (1475); Solar flares (1496); Solar coronal mass ejections(310); Solar active region magnetic fields (1975)}

\section{Introduction} \label{sec:intro}
Solar energetic particles (SEPs) are accelerated during flares and CMEs, reaching energies from Kev to GeV and consisting of protons, electrons, and ions. SEP events typically propagate along the magnetic field lines and are detected by in situ spacecraft observations. Two types of classifications can be applied to SEPs: impulsive SEP events and gradual SEP events \citep{TwoSourcesSolarEnergeticParticles}. Typically, gradual SEP events are produced in large-scale shock waves driven by CMEs, while impulsive SEP events are associated with flare acceleration \citep{ParticleaccelerationattheSun}. However, an SEP event is often associated with the onset of both a solar flare and a CME, making it difficult to distinguish the effects of the two events on SEPs.

Historically, many statistical studies have been conducted to investigate 
types of solar events which produce SEPs. For CMEs, SEPs are closely related with those having fast speed ($>900$ km s$^{-1}$) and wide angular widths ($>60^\circ$) \citep{SEPsbycmeinsolarfastregion,2008Gopalswamy}, indicating acceleration at CME-driven shocks. 
\citet{2012Park} analyzed events occurring between 1997 and 2006 and discovered that the SEP events significantly influenced by both CME speed and angular width. Specifically, the highest probability (36.1\%) was observed for full halo CMEs with speeds of at least 1500 km s$^{-1}$, whereas the lowest probability (0.9\%) was noted for partial halo CMEs with speeds ranging from 400 km s$^{-1}$ to 1000 km s$^{-1}$.
This is because the source regions of full halo CMEs are usually near the solar disk center, and the shocks they generate have a stronger geo-effectiveness towards the Earth \citep{2003gopalswamy,2007Gopalswamy}. For relation of SEPs with flares, many works show that the derived SEP occurrence probabilities increase for stronger and more westward flares \citep{kurt,BelovProtonwithX-rayflares,2015Dierckxsens}. \citet{2016papaioannou} studied SEP events that occurred 
between 1984 and 2013 and their corresponding properties. They found majority
of strong western flares (42\% of all X1.0 flares) are SEP-related but only a small percentage (0.2\%) in C-class flares. Moreover, For X-class and M-class flares, the probability of producing SEP events is serval times (2$-$4) higher when they occur in the western hemisphere compared to the eastern hemisphere \citep{2010park}. These studies have found that SEP events are associated with fast and wide CMEs, and intense flares occurring at more westward locations.

The correlations between SEPs and the properties of source regions have been studied in many works.
Most of SEP events can be traced back to the regions of ARs \citep{ReamesBook2021}. \citet{2017MichalekSunspotandAR} studied 84 SEP events suggested that the most energetic SEPs are ejected only from the associated ARs that have a large and asymmetric penumbra. \citet{2023MarroquinMcintosh} performed a statistical analysis of 181 SEPs and also found that SEPs usually
originate from the AR classified with the "k" McIntosh subclass as the second component and in a Hale class containing a $\delta$ component. 
However, there is another set of SEP events that are not associated with ARs \citep{1986Kahler,2015Gopalswamy}. These events are all accompanied by the generation of CMEs, which originate from quiescent filament and usually have weak flare signature. This suggests magnetic complexity and strong flares are not necessarily required for the production of SEPs. These studies indicate that the source regions of SEPs are not only from ARs within but also from outside of them.  

In the past, numerous studies have shown that the non-potentiality of ARs is closely related to the production of flares and CMEs. Magnetic parameters such as magnetic twists $\alpha$ \citep{1999LekaAlpha,alpha-CME2002,2006Falconer}, magnetic energy \citep{2007LekaMagenticEnergy, 2012chenParameters, 2018CuiMagenticEnergy}, magnetic helicity \citep{2004Nindoshecility}, and shear angle \citep{1984SShearHaygard,1994WangShear} typically describe the degree of non-potentiality of an AR. \cite{2016Bobra} found $\alpha$ plays an important role in discriminating between confined and eruptive flares. Recently,  
\cite{2021Gupta} found when the relative contribution of free magnetic energy, the fraction of non-potential helicity, and the normalized current-carrying helicity exceed specific thresholds, ARs are more likely to produce large CME-associated flares.
\cite{2020AvalloneSun} and \cite{2022Kazachenko} found that confined ARs tend to be more current neutralized than eruptive ARs. \cite{2022Li} proposed a new parameter describing the relative twist within the magnetic field of ARs, which effectively distinguished between eruptive and confined flares. However, the connections of the magentic parameters of the AR with SEPs have not yet been established. These parameters have previously been used to study their correlation with flare magnitude and CME speed, as well as to differentiate between eruptive and confined flares. Our aim is to ﬁrst explore the connection between magentic parameters and the occurrence of large SEP events. We find that SEP source regions have higher magnetic free energy and reconnection flux compared to No-SEP source regions. However, SEP source regions are smaller in terms of mean characteristic magnetic twist $\alpha$ and mean shear angle.

In this study, we try to categorize different SEP source regions and analyze the relations with SEPs. We consider large SEP events($>$10 MeV protons exceeding the $\geq$10 pfu threshold) recorded during 2010 to 2024. In our initial sample, there are 47 SEP events with clearly identifiable source regions, of which only two are not associated with halo-CMEs. Considering large SEPs in our study are mostly associated with halo-CMEs, we place our samples within the context of all halo-CMEs with and without SEPs to study the differences among them. This work is organized as follows: In Section \ref{sec:Data}, the process of establishing the database and calculating parameters is demonstrated. In Section \ref{sec:Results}, we present classification standard and statistical results. At last, a summary and discussion of this work is provided in Section \ref{sec:Summary and Discussion}.

\section{Data Preparation and Parameter Calculations} \label{sec:Data}

\begin{deluxetable}{cccc}
\label{Table1}
\tablecaption{Parameters Calculated to Analysis}
\tablehead{
  \colhead{Parameters} & \colhead{Description} & \colhead{Unit} & \colhead{Formula}
}
\startdata
$E_{\text{total}}$ & Total Magnetic free energy & \text{erg cm}$^{-1}$ & $E_{\text{free}} = \sum \rho_{\text{free}} dA$ \\
$J_{\text{total}}$ & Total vertical electric current & \text{A} & $J_{\text{total}} = \sum |J_z| dA$ \\
$\Phi_{\text{HED}}$ & Total unsigned magnetic flux within HED region & \text{Mx} & $\Phi_{\text{HED}} = \sum |B_z| dA$ \\
$\alpha_{\text{mean}}$ & Mean characteristic twist parameter & $\text{Mm}^{-1}$ & $\alpha_{\text{mean}}=\frac{\mu \sum J_z B_z}{\sum B_z^2}$ \\
$\Phi_{\text{AR}}$ & Total unsigned flux & $\text{Mx}$ & $\Phi_{\text{AR}} = \sum |B_z| dA$  \\
$\Psi$ & Mean shear angle & $\text{Degree}$ & $\Psi = \arccos \left( \frac{\mathbf{B}_{\text{obs}} \cdot \mathbf{B}_{\text{pot}}}{|\mathbf{B}_{\text{obs}}| |\mathbf{B}_{\text{pot}}|} \right)$\\
\enddata
\end{deluxetable}

We consider all the halo-CME events from 2010 to 2024 May, regardless of the flare class and the solar disk position. To determine if the flare is associated with a halo-CME, a catalog of halo-CMEs is used at the CDAW Data Center\footnote{\url{http://cdaw.gsfc.nasa.gov/CME_list/halo/halo.html}} \citep{2010GopalswamyHCMEsCATOLOG}, which consider only those completely surrounding the occulting disk as full halos. In this way, we compel 176 halo-CME events with identifiable source regions. In order to identify the SEPs within halo-CMEs, we use the catalog from the Coordinated Data Analysis Workshops (CDAW)\footnote{\url{https://cdaw.gsfc.nasa.gov/CME_list/sepe/}} and NOAA Space Weather Prediction Center\footnote{\url{https://www.ngdc.noaa.gov/stp/space-weather/interplanetary-data/solar-proton-events/SEP\%20page\%20code.html}}($>$10 MeV protons exceeding the $\geq$10 pfu threshold) detected by NOAA’s GOES spacecraft \citep{2002GopalswamyCDAW}. From this list, 45 large SEP events are identified out of the 176 halo-CMEs and we name it ``HCSEP''. For each event, we only consider the events whose flare and CME sources are clear. we extract the eruption information from (E)UV observations from Atmospheric Imaging Assembly (AIA, \citealp{2012Lemen}) on board the Solar Dynamics Observatory (SDO) in order to analyze their source regions.

In order to determine if the photospheric magnetic parameters can be used to distinguish SEP events and No-SEP events from halo-CMEs. The analysis includes both “intensive” parameters (those not scaling with the AR size)—such as mean characteristic magnetic twist $\alpha$ and mean shear angle, and “extensive” parameters (those scaling with the AR size). The extensive parameters include the total magnetic free energy $E_{\text{total}}$, the total vertical electric current $J_{\text{total}}$, the total unsigned magnetic flux within the  high photospheric magnetic free energy
density (HED) region $\Phi_{\text{HED}}$, and the total unsigned flux of ARs $\Phi_{\text{AR}}$. These parameters are calculated before the flare onset (within 30 min) by using the available vector magnetograms from Space-Weather Helioseismic and Magnetic Imager (HMI; \citealp{2012Scherrer})/AR Patches (SHARP; \citealp{2014Bobra}) observed by the SDO (\citealp{2012Pesnell}). Detailed information on the parameters is listed in Table \ref{Table1} \citep{2024Li}.
To reduce the impact of projection effects on the calculated parameters, we select flares from ``HCSEP'' that are greater than C5.0 and within 45° for research, and name it ``HCSEP-Sub I''. This sample consists of 72 halo-CME events, of which 25 are associated with SEPs. 

In order to analyze the relations of SEP-related events with flare parameters, we use the database ``SolarErupDB''\footnote{\url{https://solarmuri.ssl.berkeley.edu/~kazachenko/SolarErupDB/}}, which contains 481 solar flares of GOES class greater than C5.0-class and within 45° of the central meridian from 2010 to 2016 April \citep{2023Kazachenko}. In this database, we utilize their properties include $\Phi_{AR}$ (unsigned magnetic flux), $S_{AR}$ (ribbon area), $\Delta\Phi_{ribbon}$ (unsigned reconnection flux), $S_{ribbon}$ (ribbon area), $R_{\Phi}$ (reconnection flux fraction) and $R_{S}$ (ribbon area fraction). After selection, we get 37 halo-CME with 11 large SEPs from the intersection of ``SolarErupDB'' and our ``HCSEP''. We name it ``HCSEP-Sub II''. Given the limited sample size, the analysis of HCSEP-Sub II is exploratory.


\section{Results} \label{sec:Results}
\begin{figure*}
    \centering
    \includegraphics[width=0.90\textwidth]{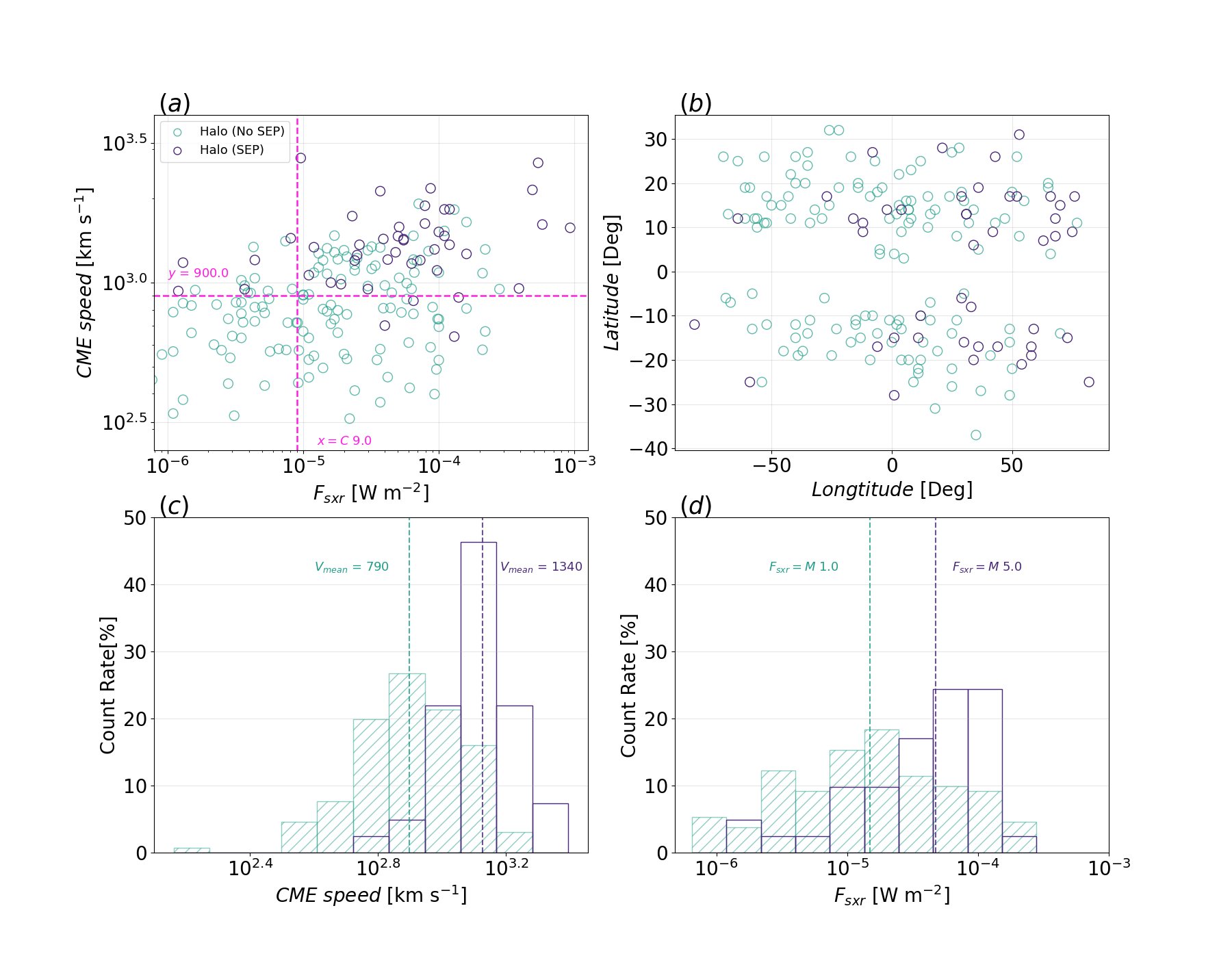}
    \caption{Overview of halo CMEs with SEPs (purple) and without SEPs (green) during 2010 Apr to 2024 May. Panel (a): Scatter plot of the CME speed as a function of GOES SXR 1–8 \AA~flux where two pink dashed lines represent 900 km s$^{-1}$ and $9\times10^{-6}$ W m$^{-2}$, respectively. Panel (b): The same as panel (a) but for flare longitude and latitude distribution on solar disk. Panel (c): The proportion histogram of speeds of the two groups of halo-CMEs. The green and purple dashed lines indicate the mean value of the two sets, respectively. Panel (d): In the same pattern as panel (c), but with the variable on the x-axis denoting GOES SXR 1–8 \AA~flux.}
    \label{fig1}
\end{figure*}

Figure \ref{fig1} shows an overview of the halo-CME events with and without SEPs from April 2010 to May 2024 for HCSEPDB. It is shown in Figure \ref{fig1} (a) that the SEP events are mostly associated with high-speed CMEs and more intense flares. More than 85\% of the SEP events exceed a CME speed of 900 km s$^{-1}$ and a flare class higher than C9.0 (represented by the pink dashed lines in Figure \ref{fig1} (a)), while those without SEPs have a lower value for both parameters statistically. Figure \ref{fig1} (b) illustrates that the source regions of SEP events are mainly located in the western hemisphere, while the events without SEPs have a more widespread distribution. In Figures \ref{fig1}(c) and (d), CME speed distribution exhibits a bimodal structure, whereas the flare flux distribution shows a large overlap for CMEs with and without SEPs. This suggests that the CME speed is a more effective indicator for distinguishing between the two groups. The mean values for both groups are represented by dashed lines in Figures \ref{fig1}(c) and (d), with SEP events corresponding to higher speeds (1340 km s$^{-1}$) for halo-CMEs and higher flare flux (M5.0).

After examining the SDO/AIA observations of the source region for HCSEPDB, we find that the corresponding source regions of the SEPs can be classified into three types: ``Single AR", ``Multiple ARs" and ``Outside of ARs". 

For "Single AR" events, two conditions are typically met. Before the flare onset, the footpoints of the filaments or flux ropes involved are located within a single AR. Additionally, during the flare, the ribbons are located at same AR.

In "Multiple ARs" events, either the footpoints of the filaments/flux ropes involved in the eruption are located in different ARs, or the flare ribbons extend across multiple ARs. If one condition is met, the event is classified as a "Multiple AR" type.

"Outside of ARs" events are related to eruptions outside AR. Among these events, the erupting large-scale filament/flux rope is located outside the ARs, lying above the neutral line between weak unipolar background fields regions and ARs. According to the classfication of the filaments, this type of filaments is categorized as ``intermediate filament" \citep{1998Engvold}.

In order to show the property of the three
types of source regions, three different events accompanied by SEPs for each type are selected as examples to be analyzed in detail (Figure \ref{fig2}). 

\subsection{Three types of source regions for SEPs} \label{subsec:3 types}

\begin{figure*}
    \centering
    \includegraphics[width=1\textwidth,height=1\textwidth]{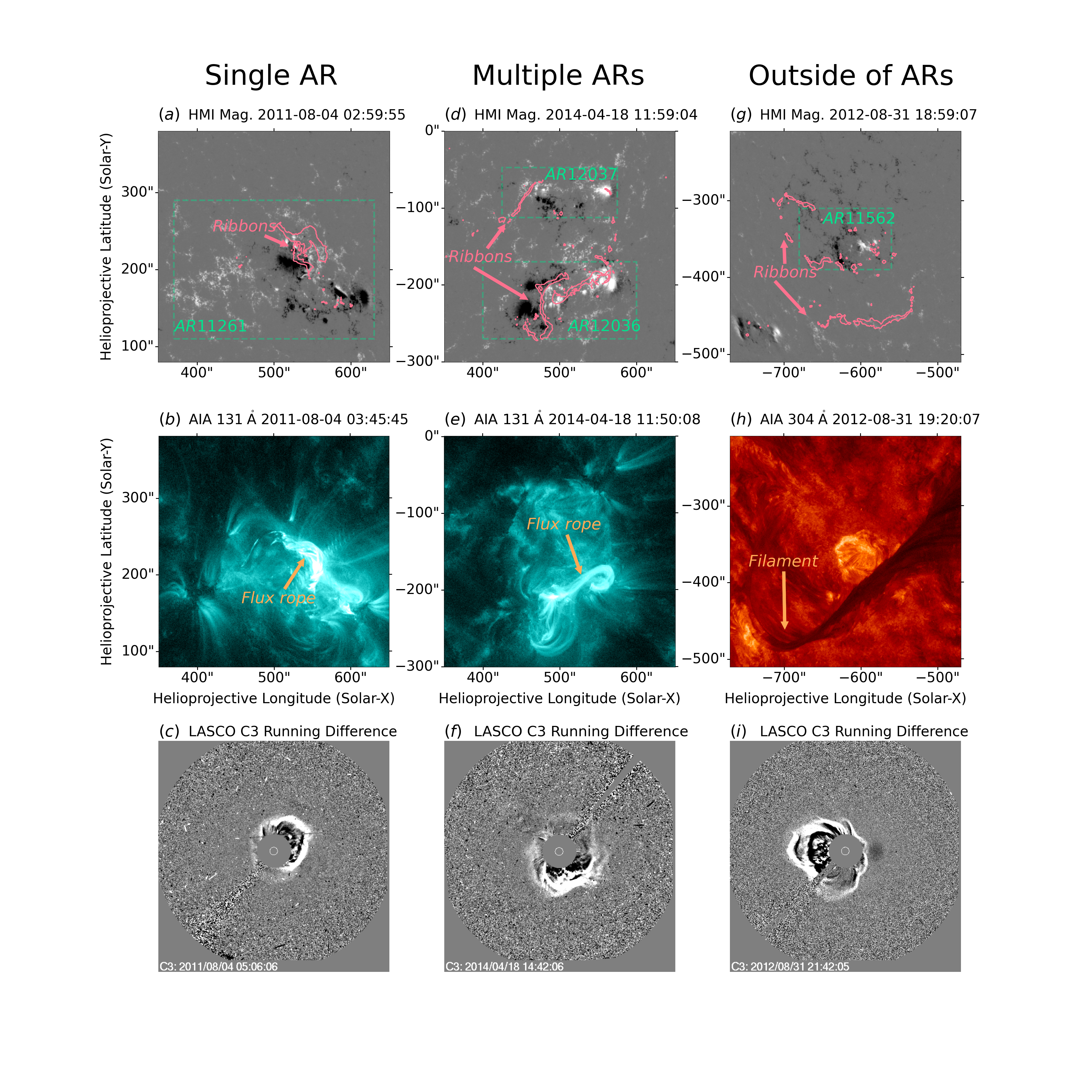}
    \caption{The HMI magnetogram, flare ribbons, EUV observations and the SOHO/LASCO C3 for three types of source regions. The top row are obtained by SDO/AIA. The flare ribbons (pink contours) are denoted on photospheric magnetograms. The flux rope and filament are observed in 131 \AA\ (b), (e) and 304 \AA\ (h). The bottom row shows running difference image from SOHO/LASCO C3 of the halo-CME associated with the flare.}
    \label{fig2}
\end{figure*}

The selected event of ``Single AR" type is the M9.3-class flare (SOL2011-08-04-T03:49) that occurred in AR 11261 on August 4, 2011. GOES SXR 1$-$8 \AA\ flux show that the flare started at 03:49 UT and peaked at 03:57 UT. Figure \ref{fig2} (a) displays that the flare ribbons are primarily located in the north of AR 11261. The northwestern part of the ribbon anchors in the positive-polarity magnetic fileds, while the southeastern part is within the negative polarity area. High-temperature flux rope is present along the neutral line between the positive and negative-polarity magnetic fields before the flare onset (Figure \ref{fig2} (b)). It subsequently forms a CME and further expands and appears as a full halo in the LASCO/C3 FOV at 05:06 UT (Figure \ref{fig2} (c)).

The selected event of ``Multiple AR" type event is the M7.3-class flare (SOL2014-04-18-T12:31) that occurred in AR 12037 and AR 12036 on 2014 April 18. GOES SXR 1$-$8 \AA\ flux showed that the flare started at 12:31 UT and peaked at 13:03 UT. The primary and secondary flare ribbons are outlined respectively in Figure \ref{fig2}(d). We observe primary flare ribbons present J-shaped are located in the lower AR 12036. An S-shaped flux rope was activated before the flare onset (Figure \ref{fig2} (e)), with the western end anchoring in the leading positive-polarity sunspot and the eastern end in the following negative polarity sunspot. The secondary flare ribbon is situated in the upper AR 12037. The CME became a halo-CME in LASCO/C3 at 14:40 UT. 

The selected event of ``Outside of AR" type is the C8.1-class flare (SOL2012-08-31-T19:45) that occurred in the region between AR 11526 and the nearby facula region on 2012 August 31. The GOES SXR 1$-$8 \AA\ flux showed that the flare started at 19:45 UT and peaked at 20:43 UT. South ribbon of this flare is located in the facula region with positive-polarity magnetic fields, while the other one spanning across the negative-polarity magetic fields of AR 11526 (Figure \ref{fig2} (g)). Figure \ref{fig2} (h) shows that the large-scale intermediate filament involved as the main body of the eruption. The magnetic neutral line corresponding to the filament is located outside the active region. The subsequent CME became a halo-CME at 21:42 (Figure \ref{fig2}(i)). 

\begin{figure*}
    \centering
    \includegraphics[width=0.80\textwidth]{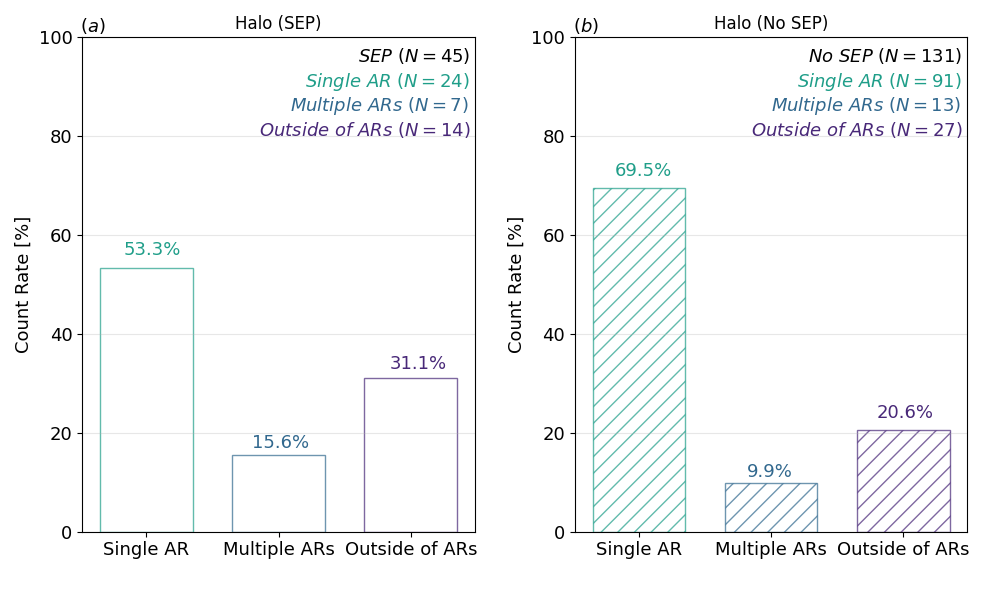}
    \caption{The proportion of three source region types for halo-CMEs with SEPs (Panel (a)) and without SEPs (Panel (b)).}
    \label{fig3}
\end{figure*}

\begin{figure*}
    \centering
    \includegraphics[width=0.90\textwidth,height=0.9\textwidth]{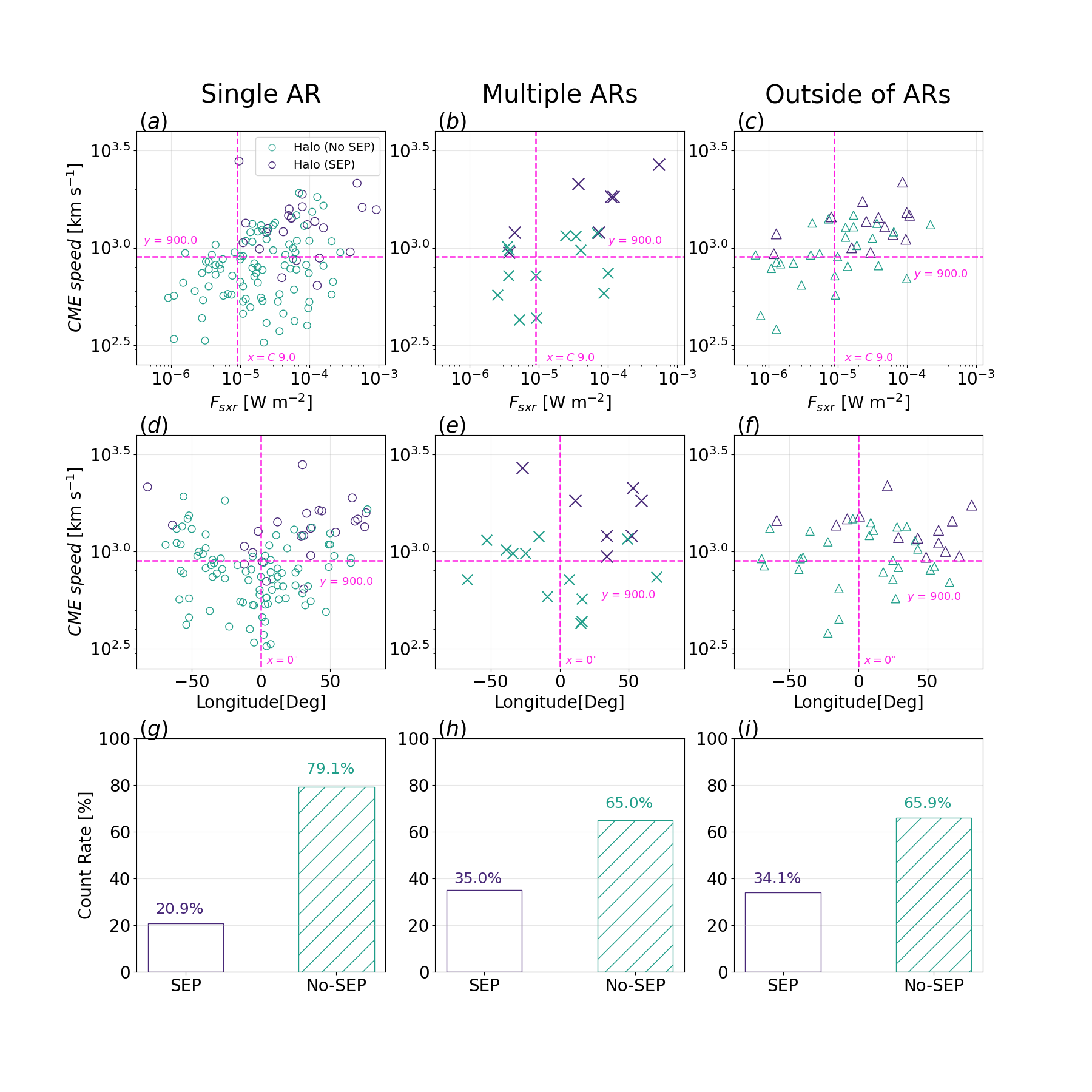}
    \caption{Statistical results of the CME speeds for SEP (purple) and No-SEP (green), GOES SXR 1$-$8 \AA\ flux, longitude distribution and SEP proportion for the three source regions. Panels (a)-(c): Scatter plots of the CME speed as a function of GOES SXR 1$-$8 \AA\ where pink dashed lines represent 900 km s$^{-1}$ and $9\times10^{-6}$ W m$^{-2}$ respectively. Panels (d)-(f): The same as (a)-(c) but with x-axis denoting longitude distribution on solar disk. The pink lines represent 900 km s$^{-1}$ and 0 $^\circ$  longitude on solar disk. Panels (g)-(i): SEP proportion in three source regions.}
    \label{fig4}
\end{figure*}

We examine the source regions for 45 SEP-related and 131 No-SEP halo-CME events, and show the statistical results in Figure \ref{fig3}. For SEP-related events, it shows that the proportion of SEP events associated with ``Single AR" is 53.3\%, 15.6\% for ``Multiple ARs" and 31.1\% for ``Outside of ARs". For No-SEP events, the proportion in ``Single AR" is 69.7\%, 9.8\% for ``Multiple ARs" and 20.5\% for ``Outside of ARs". 
We further conduct a statistical analysis of different source regions for HCSEPDB. The results are presented in Figure \ref{fig4}. In Figures \ref{fig4} (a)-(c), we find that all three source regions exhibit a trend similar to that in Figure \ref{fig1} (a), with nearly all SEPs concentrated in regions with CME speeds greater than 900 km s$^{-1}$ and flare class greater than C9.0. However, there are 5 SEP events with flare classes less than C9.0 (to the left of the pink dashed line in Figure \ref{fig4} (a)), and we find that all the 5 SEP sources belong to the ``Multiple ARs" and ``Outside of ARs" types.
In Figures \ref{fig4} (d-f), we find that fast CMEs in the western hemisphere are more likely to produce SEPs (indicated by the pink dashed lines for $>$900 km s$^{-1}$ and $>$0 $^{\circ}$ longitude). Among three source regions, the distribution of SEPs from ``Multiple ARs" is more concentrated. There are 7 events with CME speeds higher than 900 km s$^{-1}$ in western hemisphere, of which 6 are associated with SEPs (Figure \ref{fig4} (e)). But in source regions of ``Single AR'' and ``Outside of ARs'', under this criterion, only 50\% of halo-CMEs are accompanied by SEPs (Figures \ref{fig4} (d) and (f)).
The proportion of SEPs occurring in three different source regions is shown in Figures \ref{fig4} (g)-(i). We observe that the probability of SEP occurrence is similar in ``Multiple ARs" and ``Outside of ARs" types (about 35\%). But this proportion is significantly lower in the ``Single AR" (about 20\%). SEPs are more likly in ``Multiple ARs'' and ``Outside of ARs''. This indicates that SEP events tend to generate in larger source regions having greater connectivity.

\subsection{Photospheric Magnetic Parameters for SEP and No-SEP events\label{subsec:Parameters1}}
To investigate the relationship between SEP events and magnetic parameters, we use Sub I which consists of 72 halo-CME events and their photospheric magnetic parameters of source regions, 25 of which are associated with SEPs. These flares are within 45° from the central meridian to minimize the impact of projection effects on the calculated parameters. 

\begin{figure*}
    \centering
    \includegraphics[width=1\textwidth,height=0.7\textwidth]{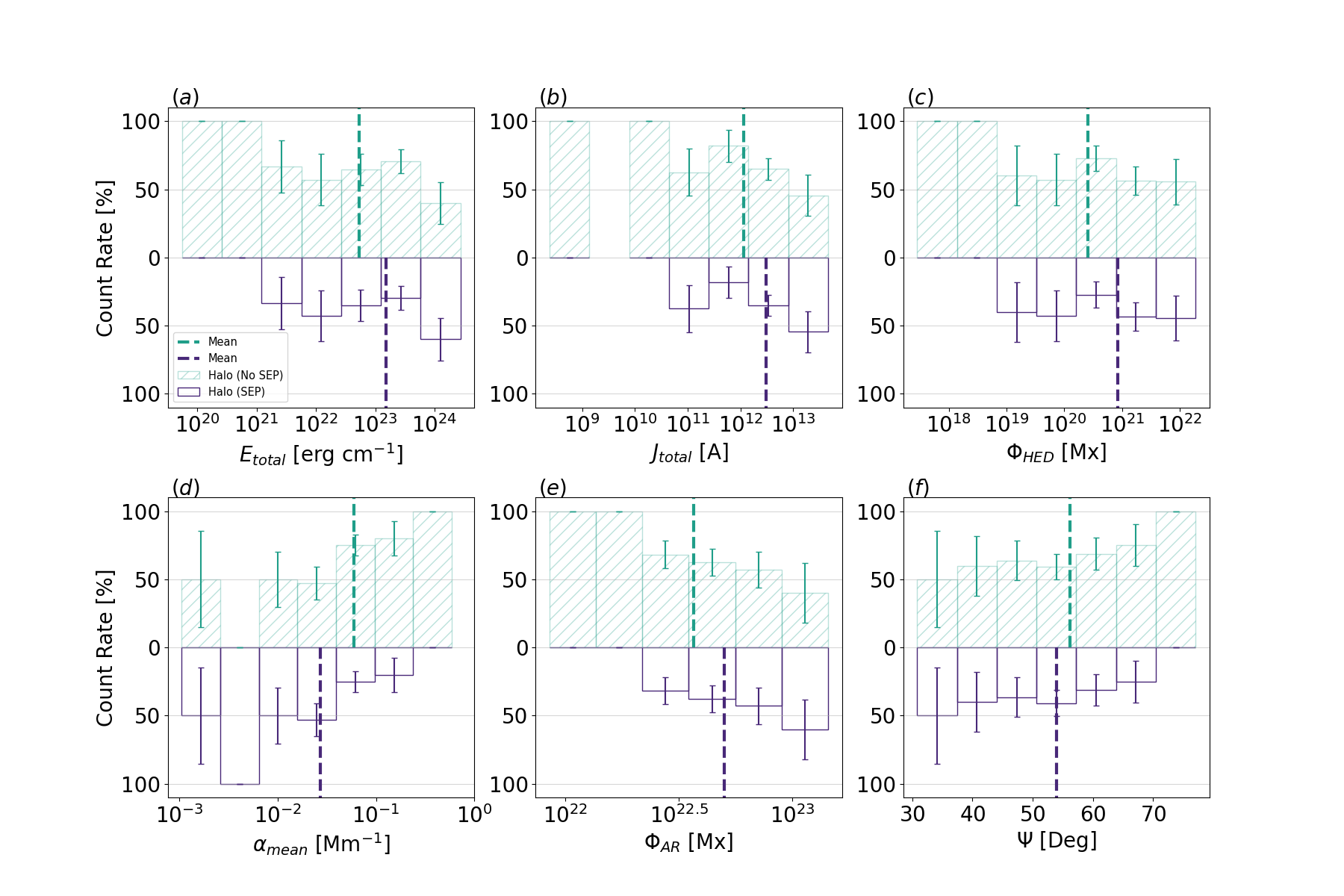}
    \caption{The percentage of SEP (purple) and No-SEP (green) events among halo-CMEs in terms of different source region magnetic parameters. The dashed lines represent the average values for SEP and No-SEP events. Panels (a)-(c): Total Magnetic free energy $E_{total}$, total vertical electric current $J_{total}$ and total unsigned magnetic flux within HED region $\Phi_{HED}$. Panels (d)-(f): Mean characteristic twist parameter $\alpha_{mean}$, Total unsigned flux $\Phi_{AR}$, Mean shear angle $\Psi$. Each bar is accompanied by error bars that indicate the standard deviation}.
    \label{fig5}
\end{figure*}

The results of the SEP and No-SEP percentage on magnetic parameters are shown in Figure \ref{fig5}. In Figure \ref{fig5} (a), (b), (c) and (e), we observe that there is no SEP occurrence in the smallest interval of $E_{total}$, $J_{total}$, $\Phi_{HED}$ and $\Phi_{AR}$. But with the continuous increase of the value, the proportion of SEP events among all halo-CMEs rises and reaches peak in the largest interval. The mean values corresponding to SEPs are also greater than those of No-SEP events in these four parameters.
However, the situation is completely reversed in $\alpha_{mean}$ and $\Psi$. The percentage of SEP events instead gradually decreases with the increase of the two parameters. Smaller $\alpha$ and $\Psi$ indicate that the twist in the photospheric magnetic field of the source region in SEP events is not strong. The percentage of SEP events originating from “Multiple ARs” and “Outside of ARs” is higher (48\%) compared to No-SEP events (Figure \ref{fig3}), which indicates that the magnetic structures of the eruptions in SEP events tend to be larger in scale. Similarly, SEP events also have higher $\Phi_{AR}$ and $\Phi_{HED}$ than that for No-SEP events. In contrast, $\alpha_{mean}$ and $\Psi$ represent the average values of HED regions \citep{2024Li}, which may be smaller under such large-scale eruptions.

\begin{figure*}
    \centering
    \includegraphics[width=1\textwidth,height=0.7\textwidth]{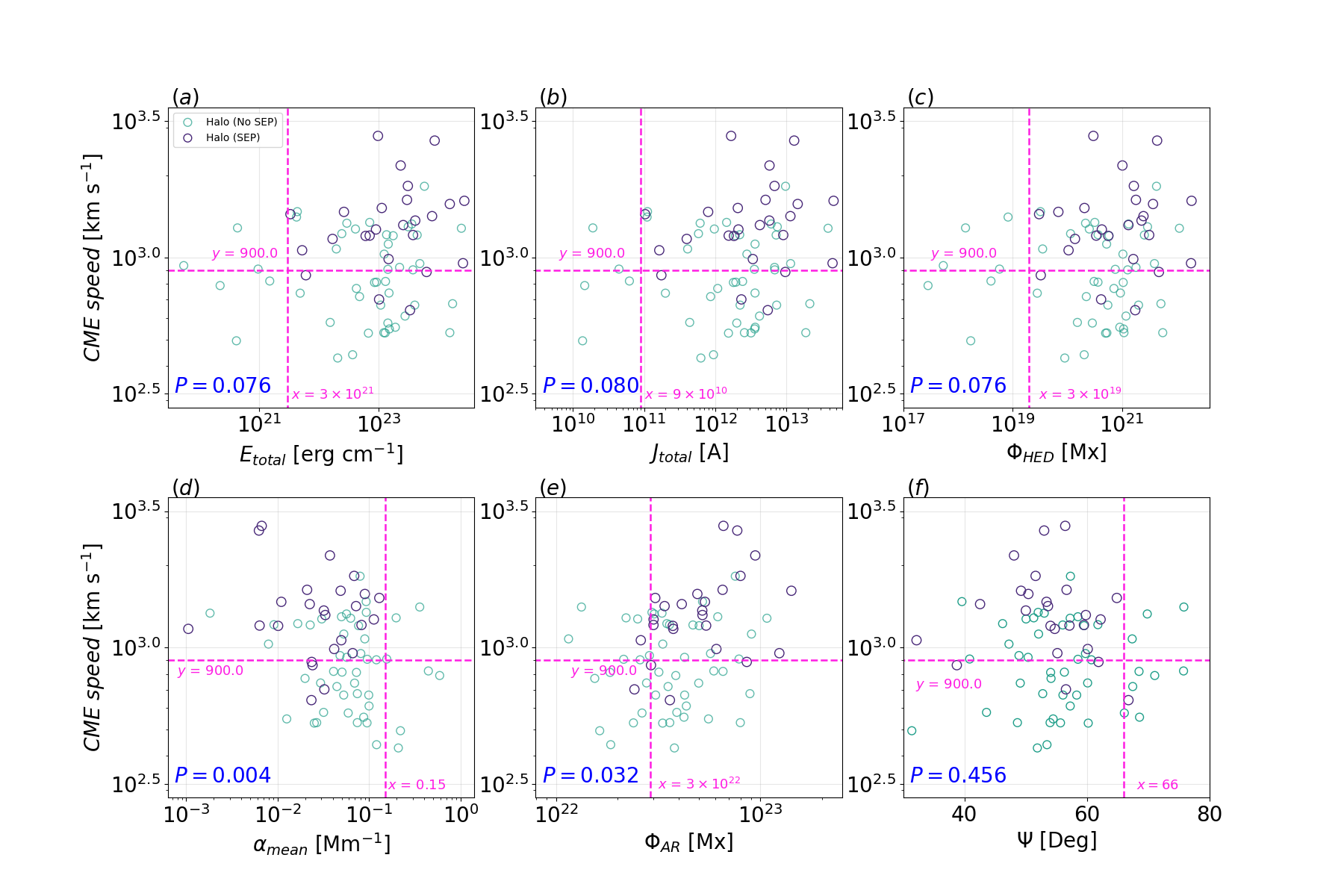}
    \caption{Scatter plots of CME speed versus different parameters for SEP (purple) and No-SEP (green). Panels (a)-(c): Total Magnetic free energy $E_{total}$, total vertical electric current $J_{total}$ and total unsigned magnetic flux within HED region $\Phi_{HED}$. Panels (d)-(f): Mean characteristic twist parameter $\alpha_{mean}$, Total unsigned flux $\Phi_{AR}$, Mean shear angle $\Psi$. Mann-Whitney U test p-values (blue) are shown to indicate significant differences.}
    \label{fig6}
\end{figure*}

To better distinguish SEP events and No-SEP events in two dimensions, we present the statistical results of the CME speeds and magnetic parameters in Figure \ref{fig6}. Overall, the CME speed shows good separation, with almost all SEP events (85\%) above the threshold of 900 km s$^{-1}$. As seen from the scatter plots in Figures \ref{fig6} (a)-(c) and (e), the distribution of SEPs is more concentrated in the upper right corner. In Figure \ref{fig6} (a), nearly 85\% SEPs have $E_{total} >3\times10^{21} $erg cm$^{-1}$ and CME speed $>$ 900 km s$^{-1}$. Similar proportion is also seen in other three panels ($J_{total} >9\times10^{10}$A, $\Phi_{HED}>3\times10^{19}$ Mx and $\Phi_{AR}>3\times10^{22}$ Mx). 
However, in Figure \ref{fig6} (d), we can observe that SEP events are concentrated in the upper left part of the graph. A majority of SEP events (21 out of 25) have a smaller magnetic twist $\alpha$ mean $<$ 0.15 (absolute value) and a larger CME speed $>$ 900 km s$^{-1}$ (indicated by the pink dashed lines). The trend is also reflected in Figure \ref{fig6} (f), with SEP events corresponding to shear angle $<66^{\circ}$ and CME speed $>$ 900 km s$^{-1}$. We performed the Mann-Whitney U test on the two datasets, and the p-values are displayed in Figure \ref{fig6}. Among all the parameters tested, only $\alpha_{mean}$ and $\Phi_{AR}$ show p-values less than 0.05. This suggests that there are significant statistical differences between SEP and No-SEP for the two parameters. $E_{total}$, $J_{total}$ and $\Phi_{HED}$ have p-values that are just above 0.05, suggesting a marginal level of significance. However, the p-value for the shear angle $\Psi$ is significantly higher than 0.05, indicating that the statistical result should not be considered reliable.
The details of the discriminating power of these parameters have been compiled into Table \ref{table2}.
\begin{table}[h]
\centering
\caption{Discriminating Power of Parameters for SEP Events}
\label{table2}
\begin{tabular}{ccc}
\toprule
Parameter 1 & Parameter 2 & SEP/ALL SEP \\
\midrule
$E_{\text{total}} > 3 \times 10^{21}$ erg cm$^{-1}$ & CME speed $> 900$ km s$^{-1}$ & 22/25 \\
$J_{\text{total}} > 9 \times 10^{10}$ A & CME speed $> 900$ km s$^{-1}$ & 22/25 \\
$\Phi_{\text{HED}} > 3 \times 10^{19}$ Mx & CME speed $> 900$ km s$^{-1}$ & 22/25 \\
$\Phi_{\text{AR}} > 3 \times 10^{22}$ Mx & CME speed $> 900$ km s$^{-1}$ & 21/25 \\
$\alpha_{\text{mean}} < 0.15$ & CME speed $> 900$ km s$^{-1}$ & 21/25 \\
$\Psi < 66^{\circ}$ & CME speed $> 900$ km s$^{-1}$ & 21/25 \\
\bottomrule
\end{tabular}
\end{table}


\subsection{Flare Ribbon Parameters for SEP and No-SEP events\label{subsec:Parameters2}}

To investigate the relationship between SEP events and flare ribbon parameters, we use Sub II which consists of 37 halo-CME events, 11 of which are associated with SEPs. These flares are within 45° and greater than C5.0.

\begin{figure*}
    \centering
    \includegraphics[width=1\textwidth,height=0.7\textwidth]{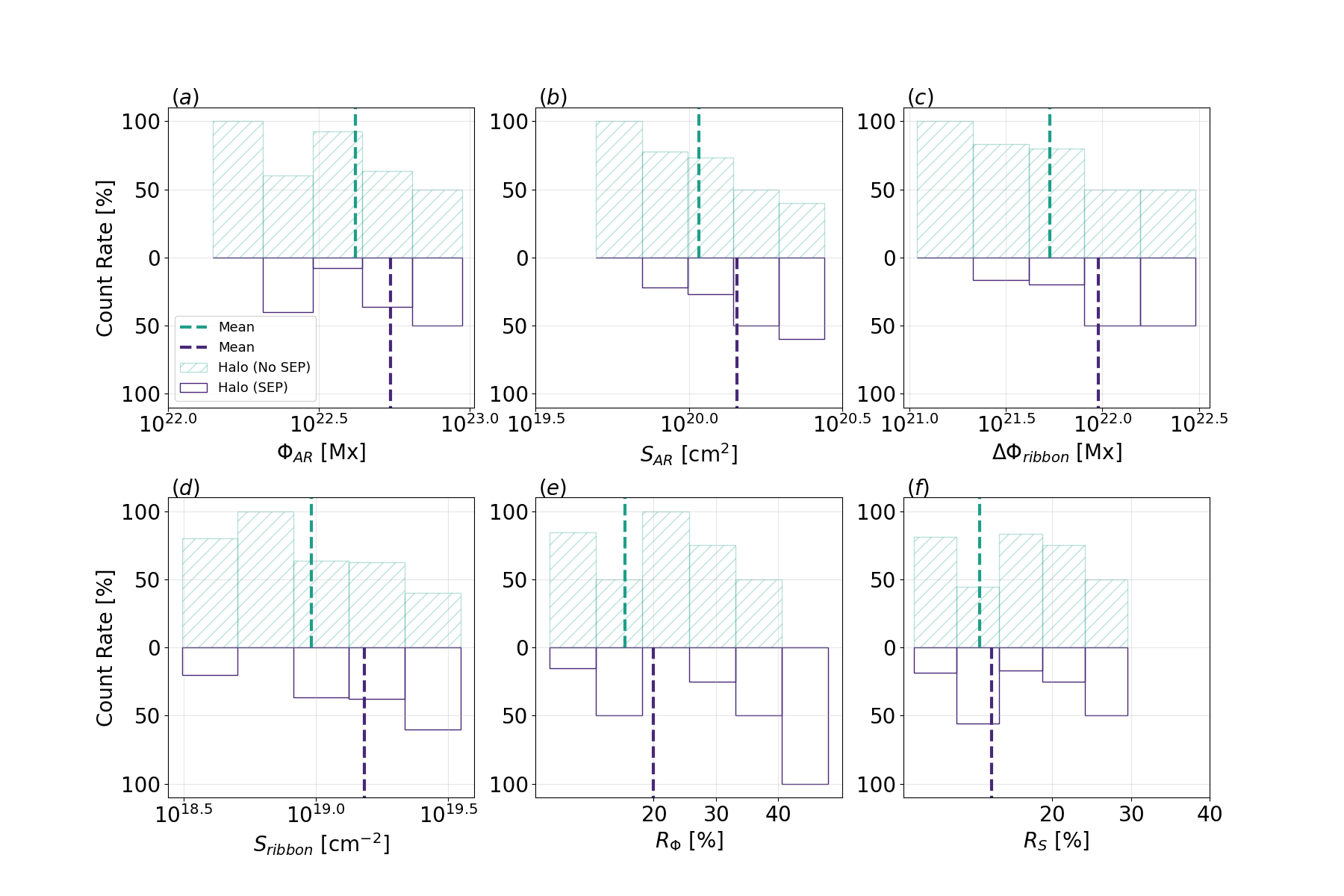}
    \caption{Similar to Figure \ref{fig5}, but in terms of diferent flare ribbon parameters of source region. Panels (a)-(c): unsigned magnetic flux $\Phi_{AR}$, ribbon area $S_{AR}$, unsigned reconnection flux $\Delta\Phi_{ribbon}$. Panels (d)-(f): ribbon area $S_{ribbon}$,  reconnection flux fraction $R_{\Phi}$ and ribbon area fraction $R_{S}$.}
    \label{fig7}
\end{figure*}

The results of the SEP and No-SEP percentage on flare ribbon parameters are shown in Figure \ref{fig7}. In distrubution of $\Phi_{AR}$, $S_{AR}$, $\Delta\Phi_{ribbon}$ and $S_{ribbon}$, we still observe that the proportion of SEPs increases with the parameter interval values (Figures \ref{fig7} (a)-(d)). However, this trend is not as significant as that of the magnetic parameters in Figure \ref{fig5}. Moreover, in $R_{\Phi}$ and $R_{S}$, there is no clear trend of increasing or decreasing SEP proportions. Overall, the distribution distinction of SEP and No-SEP events in the flare ribbon parameters (Figure \ref{fig7}) are not as pronounced as those in the magnetic parameter (Figure \ref{fig5}). The primary reason for this is the relatively small sample size of Sub I.

\begin{figure*}
    \centering
    \includegraphics[width=1\textwidth,height=0.7\textwidth]{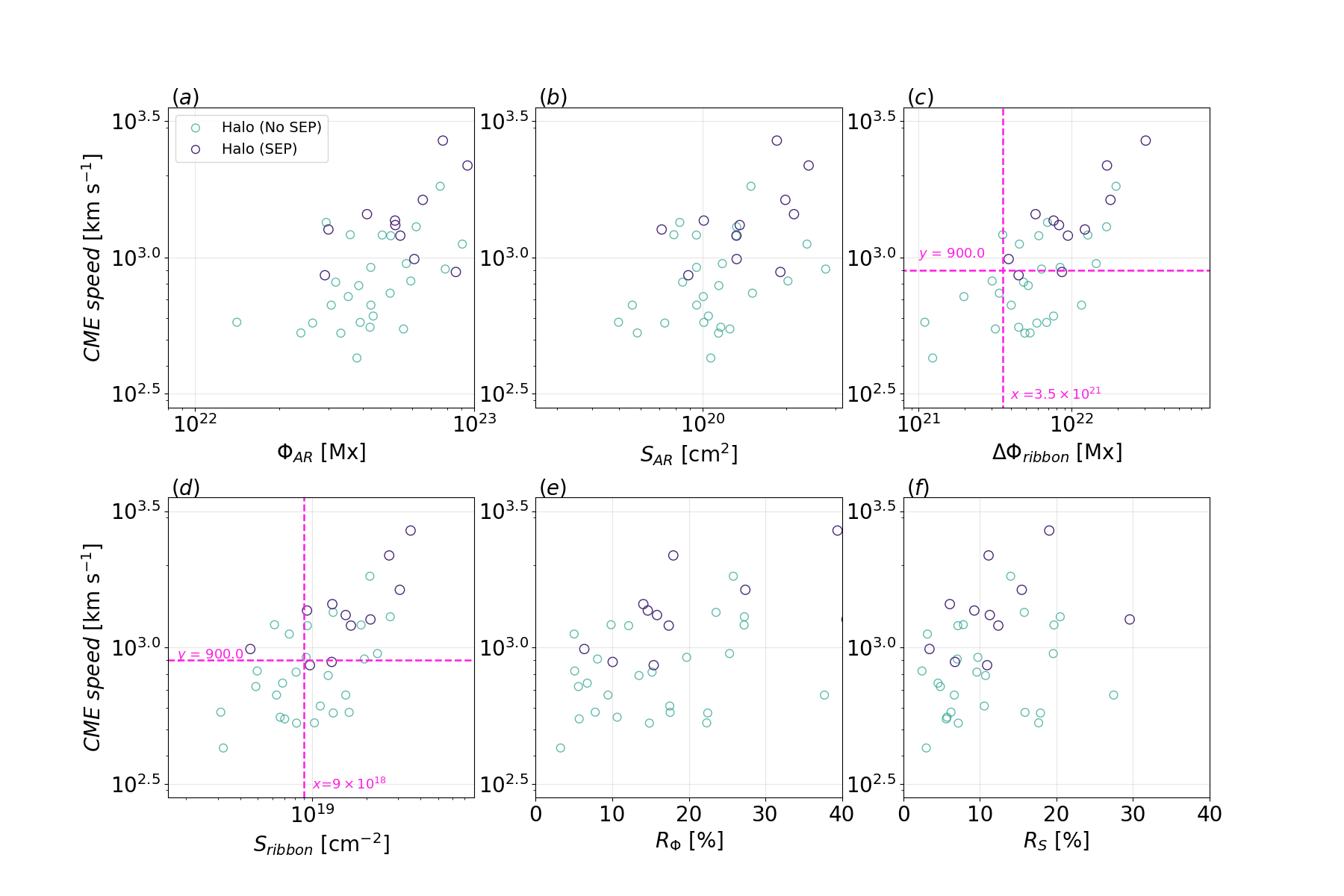}
    \caption{Scatter plots of CME speed versus flare ribbon parameters for SEP (purple) and No-SEP (green):  Panels (a)-(c): unsigned magnetic flux $\Phi_{AR}$, ribbon area $S_{AR}$, unsigned reconnection flux $\Delta\Phi_{ribbon}$. Panels (d)-(f): ribbon area $S_{ribbon}$,  reconnection flux fraction $R_{\Phi}$ and ribbon area fraction $R_{S}$.}
    \label{fig8}
\end{figure*}

Figure \ref{fig8} illustrates the CME speeds and flare ribbon parameters within Sub I.  
Overall, although the sample size has decreased, the distinction in CME speeds between SEP and No-SEP events still holds. Figures \ref{fig8} (a)-(b) show the scatter plots of $\Phi_{AR}$ and $S_{AR}$ versus CME speed. It can be seen that the distributions of the two variables are similar. The SEP events are distributed concentrating in the regions with larger values of $\Phi_{AR}$ and $S_{AR}$.  
Figures \ref{fig8} (c)-(d) show unsigned reconnection flux $\Delta\Phi_{ribbon}$ and ribbon area $S_{ribbon}$ versus CME speed, both reflecting the extent of the reconnection involved. 
In Figure \ref{fig8} (c), it can be observed that the SEPs have a concentrated distribution. 9 out of 11 CMEs have speeds larger than 900 km s$^{-1}$ and $\Delta\Phi_{ribbon}$ larger than $3.5\times10^{21}\ \text{Mx}$.
A similar distribution is also observed in Figure \ref{fig8} (d), where 8 out of 11 CMEs are faster than 900 km s$^{-1}$ and have a $S_{ribbon}$ larger than $9\times10^{18}\ \text{cm}^{-2}$ (both indicated by pink dashed lines in Figures \ref{fig8} (c)-(d)). These results indicate that the part of the AR involved in the eruption is important for producing SEPs.
In Figures \ref{fig8} (e)-(f), 
it can be seen that the distribution of the $R_{\Phi}$ and $R_{S}$ is relatively uniform for SEP and No-SEP events.

\section{Summary and Discussion} \label{sec:Summary and Discussion}

In this paper, we aim to investigate source region distinctions between halo-CMEs with and without SEP events. Therefore, we statistically analyze all 176 halo-CME with 45 SEP events from 2010 to 2024 May. We find that SEP occurrence in halo-CME strongly depends on CME speed. But the flare flux is not significantly distinguishable between the SEP events and No-SEP events in halo-CMEs.
We examine the source region characteristics of all halo-CME events and classify them into three types. We find that in SEP events, 53\% come from ``Single AR'', while the rest ones originating from ``Multiple ARs'' or ``Outside of ARs''. In contrast, in No-SEP events, the proportion is higher in ``Single AR'' (30\%) and lower in ``Multiple ARs'' or ``Outside of ARs'' (70\%). 
We further analyze the distribution of SEP and No-SEP events in different source regions. In ``Multiple ARs'', among all halo-CMEs occurring in the western hemisphere with CME speeds exceeding 900 km s$^{-1}$, 90\% are associated with SEPs. This indicates that in ``Multiple ARs'', CME speed and source region location are useful indicators for forecasting SEPs.
We also calculate the magnetic parameters of the source regions. Compared to No-SEP events, SEP source regions have higher magnetic flux and free energy but have lower magnetic twist $\alpha$ and shear angles.

From the distribution of SEP and No-SEP events among the overall halo-CME events, we find that the CME speed is a key parameter in distinguishing SEP and No-SEP events. The CME speeds corresponding to almost all SEP events are above 900 km s$^{-1}$, which is similar to the conclusions of \citet{SEPsbycmeinsolarfastregion} and \citet{2008Gopalswamy}. There is a distinct bimodal structure for both SEP and No-SEP events in the histogram of CME speeds, with the former having a much higher average value (v=1340 $\text{km}\ \text{s}^{-1}$) than the latter (v=790 $\text{km}\ \text{s}^{-1}$). CME speeds associated with SEP events are higher than the average speed of halo-CMEs (v=998 $\text{km}\ \text{s}^{-1}$) \citep{2010Gopalswamy}, which indicates that SEP-producing CMEs need to have above-average speed to drive a shock. However, flare flux does not effectively differentiate the two groups and the distribution of the histogram shows a certain degree of overlap. It can be seen that there is no obvious bimodal structure. Our results
show that the SEP events is closely related to fast CMEs, while the relationship with flare flux is not significant. In our analysis, all SEP events are large SEP events($>$10 MeV protons exceeding the $\geq$10 pfu threshold). Large SEP events are 
thought to be primarily due to shock acceleration \citep{ParticleaccelerationattheSun}, which is consistent with our results. In the past, 
\cite{2003Gopalswamy-} reported that the occurrence rates of large SEP events, DH type II bursts, and fast ($>900$ km s$^{-1}$) and wide ($>60^\circ$) frontside western hemispheric CMEs are highly consistent. This supports that CME-driven shocks can accelerate both protons and electrons.
\cite{2012Park} also found that the large SEPs probabilities with fast CMEs ($>1500$ km s$^{-1}$) is the highest (32\%), and the correlation coefficient between SEP peak flux and CME speed (0.69) is higher than that with flare flux (0.30). These studies also support that large SEPs are closely related to fast CMEs. 

In previous studies, CME source regions are typically associated with filament eruptions in ARs and quiet-Sun regions \citep{2001Subramanian}. However, in our statistical analysis, we find that many AR source regions are not dominated by a single AR but involved the combination of multiple ARs. In our work we reclassify the AR source regions based on their scale into ``Single AR" and ``Multiple AR" categories. The ``PEs" category in \citet{2001Subramanian} is similar to our "Outside of ARs" type, both indicating CME source regions related to eruptions outside ARs.
Our work also show the differences in source region between SEP events and No-SEP events. We found 53\% of the SEP events in halo-CMEs correspond to ``Single AR", 16\% are associated with ``Multiple ARs", and 31\% occur ``Outside of ARs". For No-SEP events, the proportion are 70\% , 9\% and 21\%. The source regions of SEP events have a higher proportion in the ``Multiple ARs" and ``Outside of ARs" categories compared to No-SEP events. 
The characteristics of these two types of source regions indicates that CMEs are intrinsically large-scale events, wherein their energy supply, initiation mechanisms, and final angular widths are all derived from large-scale regions \citep{2006Zhou,2008Van}.
\cite{2015Gopalswamy} reported four large filament eruptions outside ARs causing SEP events. These four large filaments all span up to more than 400 arcsecs on the disk, suggesting that they are inherently nature large-scale events. The ``Multiple ARs'' involve the participation of complicated magnetic topology and the significant characteristic being the extension of secondary ribbons. This may result from large
scale secondary magnetic reconnection triggered by the initial
 magnetic reconnection in the main flare regions \citep{2011Woods,2013Liu,2013Sun}. This indicates source region of a CME extends well beyond the initiation site and is large scale by nature. The higher percentage of SEP events in these two types of source regions indicates that their source regions are more widely distributed at nature large-scale.
 
We have also discovered some patterns in the parameter statistics of SEP and No-SEP events. We observe that SEP events tend to have higher values in $E_{total}$, $J_{total}$, $\Phi_{HED}$ and $\Phi_{AR}$. SEP events are also concentrated in source regions with larger $\Delta\Phi_{ribbon}$ and $S_{ribbon}$. This suggests that regions with higher free magnetic energy and stronger reconnection magnetic flux are more likely to produce SEPs. We also find that SEP events correspond to smaller $\alpha$ values and shear angles than No-SEP events. Approximately 85\% of SEP events have $\alpha <$0.15 and CME speed$>$900 km s$^{-1}$ (or $\Psi<$66$^{\circ}$ and CME speed$>$900 km s$^{-1}$), while the proportion in No-SEPs is about 50\%. There may be two reasons for this. First, we find SEP source regions are larger in scale. This indicates that the eruption takes place in higher layers of the solar atmosphere. As a result, the filaments and flux ropes in the higher corona may not necessarily manifest as strong twist in the photosphere. 
Second, we find about 90\% SEP events originate from ARs with $\Phi_{AR}>$ 3$\times 10^{22}$ Mx. In the statistical results of \citet{2022Li}, the mean characteristic parameter $\alpha$ and shear angle show anti-correlation with the $\Phi_{AR}$ (Figure 4 in \citealp{2022Li}). This implies that larger ARs tend to have smaller $\alpha$ value and $\Psi$. The SEP events tend to originate from larger ARs, and thus have relatively smaller $\alpha$ and $\Psi$. These values are mean values calculated within the high free energy region, which does not represent the total twist for the filament or flux rope. Our results suggest that SEP events originate from regions with higher extensive properties and lower intensive properties.

Our research indicates that the occurrence of SEPs in halo-CMEs is related to large-scale eruptive source regions, while those of No-SEPs are relatively smaller. Moreover, SEP events show greater involvement of free energy and magnetic flux. However, SEP source regions demonstrate small values in intensive properties before eruption. A deeper understanding of the mechanisms behind SEP events is still needed.

\section*{Acknowledgments}
We thank Doctor Jing-Jing Wang at National Space Science Center for the precious discussion and suggestions of this work. The CME catalog employed here is generated and maintained at the CDAW Data Center by NASA and the Catholic University of America in cooperation with the Naval Research Laboratory. We acknowledge the use of data from NOAA-GOES missions and thank the team for the availability of particle data. SOHO
is a project of international cooperation between ESA and NASA. We also acknowledge the SDO team for making the AIA data
available. SDO is a mission for NASA's Living With a Star (LWS) program. This work is supported by the B-type Strategic Priority Program of the Chinese Academy of Sciences (grant No. XDB0560000), the National Natural Science Foundations of China (12222306, 42474220, 12273060), the National Key R\&D Programs of China (2022YFF0503800), and the Youth Innovation Promotion Association of CAS (2023063).

\bibliography{main}{}

\begin{thebibliography}{}
\expandafter\ifx\csname natexlab\endcsname\relax\def\natexlab#1{#1}\fi
\providecommand{\url}[1]{\href{#1}{#1}}
\providecommand{\dodoi}[1]{doi:~\href{http://doi.org/#1}{\nolinkurl{#1}}}
\providecommand{\doeprint}[1]{\href{http://ascl.net/#1}{\nolinkurl{http://ascl.net/#1}}}
\providecommand{\doarXiv}[1]{\href{https://arxiv.org/abs/#1}{\nolinkurl{https://arxiv.org/abs/#1}}}

\bibitem[{{Avallone} \& {Sun}(2020)}]{2020AvalloneSun}
{Avallone}, E.~A., \& {Sun}, X. 2020, \apj, 893, 123, \dodoi{10.3847/1538-4357/ab7afa}

\bibitem[{{Belov} {et~al.}(2005){Belov}, {Garcia}, {Kurt}, {Mavromichalaki}, \& {Gerontidou}}]{BelovProtonwithX-rayflares}
{Belov}, A., {Garcia}, H., {Kurt}, V., {Mavromichalaki}, H., \& {Gerontidou}, M. 2005, \solphys, 229, 135, \dodoi{10.1007/s11207-005-4721-3}

\bibitem[{{Bobra} \& {Ilonidis}(2016)}]{2016Bobra}
{Bobra}, M.~G., \& {Ilonidis}, S. 2016, \apj, 821, 127, \dodoi{10.3847/0004-637X/821/2/127}

\bibitem[{{Bobra} {et~al.}(2014){Bobra}, {Sun}, {Hoeksema}, {Turmon}, {Liu}, {Hayashi}, {Barnes}, \& {Leka}}]{2014Bobra}
{Bobra}, M.~G., {Sun}, X., {Hoeksema}, J.~T., {et~al.} 2014, \solphys, 289, 3549, \dodoi{10.1007/s11207-014-0529-3}

\bibitem[{{Bronarska} \& {Michalek}(2017)}]{2017MichalekSunspotandAR}
{Bronarska}, K., \& {Michalek}, G. 2017, Advances in Space Research, 59, 384, \dodoi{10.1016/j.asr.2016.09.011}

\bibitem[{{Chen} \& {Wang}(2012)}]{2012chenParameters}
{Chen}, A.~Q., \& {Wang}, J.~X. 2012, \aap, 543, A49, \dodoi{10.1051/0004-6361/201118037}

\bibitem[{{Cui} {et~al.}(2018){Cui}, {Wang}, {Xu}, \& {Liu}}]{2018CuiMagenticEnergy}
{Cui}, Y., {Wang}, H., {Xu}, Y., \& {Liu}, S. 2018, Journal of Geophysical Research (Space Physics), 123, 1704, \dodoi{10.1002/2017JA024710}

\bibitem[{{Dierckxsens} {et~al.}(2015){Dierckxsens}, {Tziotziou}, {Dalla}, {Patsou}, {Marsh}, {Crosby}, {Malandraki}, \& {Tsiropoula}}]{2015Dierckxsens}
{Dierckxsens}, M., {Tziotziou}, K., {Dalla}, S., {et~al.} 2015, \solphys, 290, 841, \dodoi{10.1007/s11207-014-0641-4}

\bibitem[{{Engvold}(1998)}]{1998Engvold}
{Engvold}, O. 1998, in Astronomical Society of the Pacific Conference Series, Vol. 150, IAU Colloq. 167: New Perspectives on Solar Prominences, ed. D.~F. {Webb}, B.~{Schmieder}, \& D.~M. {Rust}, 23

\bibitem[{{Falconer} {et~al.}(2002){Falconer}, {Moore}, \& {Gary}}]{alpha-CME2002}
{Falconer}, D.~A., {Moore}, R.~L., \& {Gary}, G.~A. 2002, \apj, 569, 1016, \dodoi{10.1086/339161}

\bibitem[{{Falconer} {et~al.}(2006){Falconer}, {Moore}, \& {Gary}}]{2006Falconer}
---. 2006, \apj, 644, 1258, \dodoi{10.1086/503699}

\bibitem[{{Gopalswamy} {et~al.}(2003{\natexlab{a}}){Gopalswamy}, {Lara}, {Yashiro}, {Nunes}, \& {Howard}}]{2003gopalswamy}
{Gopalswamy}, N., {Lara}, A., {Yashiro}, S., {Nunes}, S., \& {Howard}, R.~A. 2003{\natexlab{a}}, in ESA Special Publication, Vol. 535, Solar Variability as an Input to the Earth's Environment, ed. A.~{Wilson}, 403--414

\bibitem[{{Gopalswamy} {et~al.}(2015){Gopalswamy}, {M{\"a}kel{\"a}}, {Akiyama}, {Yashiro}, {Xie}, {Thakur}, \& {Kahler}}]{2015Gopalswamy}
{Gopalswamy}, N., {M{\"a}kel{\"a}}, P., {Akiyama}, S., {et~al.} 2015, \apj, 806, 8, \dodoi{10.1088/0004-637X/806/1/8}

\bibitem[{{Gopalswamy} {et~al.}(2007){Gopalswamy}, {Yashiro}, \& {Akiyama}}]{2007Gopalswamy}
{Gopalswamy}, N., {Yashiro}, S., \& {Akiyama}, S. 2007, Journal of Geophysical Research (Space Physics), 112, A06112, \dodoi{10.1029/2006JA012149}

\bibitem[{{Gopalswamy} {et~al.}(2008){Gopalswamy}, {Yashiro}, {Akiyama}, {M{\"a}kel{\"a}}, {Xie}, {Kaiser}, {Howard}, \& {Bougeret}}]{2008Gopalswamy}
{Gopalswamy}, N., {Yashiro}, S., {Akiyama}, S., {et~al.} 2008, Annales Geophysicae, 26, 3033, \dodoi{10.5194/angeo-26-3033-2008}

\bibitem[{{Gopalswamy} {et~al.}(2003{\natexlab{b}}){Gopalswamy}, {Yashiro}, {Lara}, {Kaiser}, {Thompson}, {Gallagher}, \& {Howard}}]{2003Gopalswamy-}
{Gopalswamy}, N., {Yashiro}, S., {Lara}, A., {et~al.} 2003{\natexlab{b}}, \grl, 30, 8015, \dodoi{10.1029/2002GL016435}

\bibitem[{{Gopalswamy} {et~al.}(2002){Gopalswamy}, {Yashiro}, {Micha{\l}ek}, {Kaiser}, {Howard}, {Reames}, {Leske}, \& {von Rosenvinge}}]{2002GopalswamyCDAW}
{Gopalswamy}, N., {Yashiro}, S., {Micha{\l}ek}, G., {et~al.} 2002, \apjl, 572, L103, \dodoi{10.1086/341601}

\bibitem[{{Gopalswamy} {et~al.}(2010{\natexlab{a}}){Gopalswamy}, {Yashiro}, {Michalek}, {Xie}, {M{\"a}kel{\"a}}, {Vourlidas}, \& {Howard}}]{2010GopalswamyHCMEsCATOLOG}
{Gopalswamy}, N., {Yashiro}, S., {Michalek}, G., {et~al.} 2010{\natexlab{a}}, Sun and Geosphere, 5, 7

\bibitem[{{Gopalswamy} {et~al.}(2010{\natexlab{b}}){Gopalswamy}, {Yashiro}, {Michalek}, {Xie}, {M{\"a}kel{\"a}}, {Vourlidas}, \& {Howard}}]{2010Gopalswamy}
---. 2010{\natexlab{b}}, Sun and Geosphere, 5, 7

\bibitem[{{Gupta} {et~al.}(2021){Gupta}, {Thalmann}, \& {Veronig}}]{2021Gupta}
{Gupta}, M., {Thalmann}, J.~K., \& {Veronig}, A.~M. 2021, \aap, 653, A69, \dodoi{10.1051/0004-6361/202140591}

\bibitem[{{Hagyard} {et~al.}(1984){Hagyard}, {Smith}, {Teuber}, \& {West}}]{1984SShearHaygard}
{Hagyard}, M.~J., {Smith}, Jr., J.~B., {Teuber}, D., \& {West}, E.~A. 1984, \solphys, 91, 115, \dodoi{10.1007/BF00213618}

\bibitem[{{Kahler} {et~al.}(1986){Kahler}, {Cliver}, {Cane}, {McGuire}, {Stone}, \& {Sheeley}}]{1986Kahler}
{Kahler}, S.~W., {Cliver}, E.~W., {Cane}, H.~V., {et~al.} 1986, \apj, 302, 504, \dodoi{10.1086/164009}

\bibitem[{{Kahler} \& {Reames}(2003)}]{SEPsbycmeinsolarfastregion}
{Kahler}, S.~W., \& {Reames}, D.~V. 2003, \apj, 584, 1063, \dodoi{10.1086/345780}

\bibitem[{{Kazachenko}(2023)}]{2023Kazachenko}
{Kazachenko}, M.~D. 2023, \apj, 958, 104, \dodoi{10.3847/1538-4357/ad004e}

\bibitem[{{Kazachenko} {et~al.}(2022){Kazachenko}, {Lynch}, {Savcheva}, {Sun}, \& {Welsch}}]{2022Kazachenko}
{Kazachenko}, M.~D., {Lynch}, B.~J., {Savcheva}, A., {Sun}, X., \& {Welsch}, B.~T. 2022, \apj, 926, 56, \dodoi{10.3847/1538-4357/ac3af3}

\bibitem[{{Kurt} {et~al.}(2004){Kurt}, {Belov}, {Mavromichalaki}, \& {Gerontidou}}]{kurt}
{Kurt}, V., {Belov}, A., {Mavromichalaki}, H., \& {Gerontidou}, M. 2004, Annales Geophysicae, 22, 2255, \dodoi{10.5194/angeo-22-2255-2004}

\bibitem[{{Leka} \& {Barnes}(2007)}]{2007LekaMagenticEnergy}
{Leka}, K.~D., \& {Barnes}, G. 2007, \apj, 656, 1173, \dodoi{10.1086/510282}

\bibitem[{{Leka} \& {Skumanich}(1999)}]{1999LekaAlpha}
{Leka}, K.~D., \& {Skumanich}, A. 1999, \solphys, 188, 3, \dodoi{10.1023/A:1005108632671}

\bibitem[{{Lemen} {et~al.}(2012){Lemen}, {Title}, {Akin}, {Boerner}, {Chou}, {Drake}, {Duncan}, {Edwards}, {Friedlaender}, {Heyman}, {Hurlburt}, {Katz}, {Kushner}, {Levay}, {Lindgren}, {Mathur}, {McFeaters}, {Mitchell}, {Rehse}, {Schrijver}, {Springer}, {Stern}, {Tarbell}, {Wuelser}, {Wolfson}, {Yanari}, {Bookbinder}, {Cheimets}, {Caldwell}, {Deluca}, {Gates}, {Golub}, {Park}, {Podgorski}, {Bush}, {Scherrer}, {Gummin}, {Smith}, {Auker}, {Jerram}, {Pool}, {Soufli}, {Windt}, {Beardsley}, {Clapp}, {Lang}, \& {Waltham}}]{2012Lemen}
{Lemen}, J.~R., {Title}, A.~M., {Akin}, D.~J., {et~al.} 2012, \solphys, 275, 17, \dodoi{10.1007/s11207-011-9776-8}

\bibitem[{{Li} {et~al.}(2022){Li}, {Sun}, {Hou}, {Chen}, {Yang}, \& {Zhang}}]{2022Li}
{Li}, T., {Sun}, X., {Hou}, Y., {et~al.} 2022, \apjl, 926, L14, \dodoi{10.3847/2041-8213/ac5251}

\bibitem[{{Li} {et~al.}(2024){Li}, {Zheng}, {Li}, {Hou}, {Li}, {Zhang}, \& {Chen}}]{2024Li}
{Li}, T., {Zheng}, Y., {Li}, X., {et~al.} 2024, \apj, 964, 159, \dodoi{10.3847/1538-4357/ad2e90}

\bibitem[{{Liu} {et~al.}(2013){Liu}, {Zhang}, {Wang}, \& {Cheng}}]{2013Liu}
{Liu}, K., {Zhang}, J., {Wang}, Y., \& {Cheng}, X. 2013, \apj, 768, 150, \dodoi{10.1088/0004-637X/768/2/150}

\bibitem[{{Marroquin} {et~al.}(2023){Marroquin}, {Sadykov}, {Kosovichev}, {Kitiashvili}, {Oria}, {Nita}, {Illarionov}, {O'Keefe}, {Francis}, {Chong}, {Kosovich}, \& {Ali}}]{2023MarroquinMcintosh}
{Marroquin}, R.~D., {Sadykov}, V., {Kosovichev}, A., {et~al.} 2023, \apj, 952, 97, \dodoi{10.3847/1538-4357/acdb65}

\bibitem[{{Nindos} \& {Andrews}(2004)}]{2004Nindoshecility}
{Nindos}, A., \& {Andrews}, M.~D. 2004, \apjl, 616, L175, \dodoi{10.1086/426861}

\bibitem[{{Papaioannou} {et~al.}(2016){Papaioannou}, {Sandberg}, {Anastasiadis}, {Kouloumvakos}, {Georgoulis}, {Tziotziou}, {Tsiropoula}, {Jiggens}, \& {Hilgers}}]{2016papaioannou}
{Papaioannou}, A., {Sandberg}, I., {Anastasiadis}, A., {et~al.} 2016, Journal of Space Weather and Space Climate, 6, A42, \dodoi{10.1051/swsc/2016035}

\bibitem[{{Park} {et~al.}(2012){Park}, {Moon}, \& {Gopalswamy}}]{2012Park}
{Park}, J., {Moon}, Y.~J., \& {Gopalswamy}, N. 2012, Journal of Geophysical Research (Space Physics), 117, A08108, \dodoi{10.1029/2011JA017477}

\bibitem[{{Park} {et~al.}(2010){Park}, {Moon}, {Lee}, \& {Youn}}]{2010park}
{Park}, J., {Moon}, Y.-J., {Lee}, D.-H., \& {Youn}, S. 2010, in 38th COSPAR Scientific Assembly, Vol.~38, 5

\bibitem[{{Pesnell} {et~al.}(2012){Pesnell}, {Thompson}, \& {Chamberlin}}]{2012Pesnell}
{Pesnell}, W.~D., {Thompson}, B.~J., \& {Chamberlin}, P.~C. 2012, \solphys, 275, 3, \dodoi{10.1007/s11207-011-9841-3}

\bibitem[{{Reames}(1999)}]{ParticleaccelerationattheSun}
{Reames}, D.~V. 1999, \ssr, 90, 413, \dodoi{10.1023/A:1005105831781}

\bibitem[{{Reames}(2013)}]{TwoSourcesSolarEnergeticParticles}
---. 2013, \ssr, 175, 53, \dodoi{10.1007/s11214-013-9958-9}

\bibitem[{{Reames}(2021)}]{ReamesBook2021}
---. 2021, {Solar Energetic Particles. A Modern Primer on Understanding Sources, Acceleration and Propagation}, Vol. 978, \dodoi{10.1007/978-3-030-66402-2}

\bibitem[{{Scherrer} {et~al.}(2012){Scherrer}, {Schou}, {Bush}, {Kosovichev}, {Bogart}, {Hoeksema}, {Liu}, {Duvall}, {Zhao}, {Title}, {Schrijver}, {Tarbell}, \& {Tomczyk}}]{2012Scherrer}
{Scherrer}, P.~H., {Schou}, J., {Bush}, R.~I., {et~al.} 2012, \solphys, 275, 207, \dodoi{10.1007/s11207-011-9834-2}

\bibitem[{{Subramanian} \& {Dere}(2001)}]{2001Subramanian}
{Subramanian}, P., \& {Dere}, K.~P. 2001, \apj, 561, 372, \dodoi{10.1086/323213}

\bibitem[{{Sun} {et~al.}(2013){Sun}, {Hoeksema}, {Liu}, {Aulanier}, {Su}, {Hannah}, \& {Hock}}]{2013Sun}
{Sun}, X., {Hoeksema}, J.~T., {Liu}, Y., {et~al.} 2013, \apj, 778, 139, \dodoi{10.1088/0004-637X/778/2/139}

\bibitem[{van Driel-Gesztelyi {et~al.}(2008)van Driel-Gesztelyi, Attrill, D\'emoulin, Mandrini, \& Harra}]{2008Van}
van Driel-Gesztelyi, L., Attrill, G. D.~R., D\'emoulin, P., Mandrini, C.~H., \& Harra, L.~K. 2008, Annales Geophysicae, 26, 3077, \dodoi{10.5194/angeo-26-3077-2008}

\bibitem[{{Wang} {et~al.}(1994){Wang}, {Ewell}, {Zirin}, \& {Ai}}]{1994WangShear}
{Wang}, H., {Ewell}, Jr., M.~W., {Zirin}, H., \& {Ai}, G. 1994, \apj, 424, 436, \dodoi{10.1086/173901}

\bibitem[{{Woods} {et~al.}(2011){Woods}, {Hock}, {Eparvier}, {Jones}, {Chamberlin}, {Klimchuk}, {Didkovsky}, {Judge}, {Mariska}, {Warren}, {Schrijver}, {Webb}, {Bailey}, \& {Tobiska}}]{2011Woods}
{Woods}, T.~N., {Hock}, R., {Eparvier}, F., {et~al.} 2011, \apj, 739, 59, \dodoi{10.1088/0004-637X/739/2/59}

\bibitem[{{Zhou} {et~al.}(2006){Zhou}, {Wang}, \& {Zhang}}]{2006Zhou}
{Zhou}, G.~P., {Wang}, J.~X., \& {Zhang}, J. 2006, \aap, 445, 1133, \dodoi{10.1051/0004-6361:20053536}

\end{thebibliography}
\bibliographystyle{aasjournal}

\end{document}